# What's next? Forecasting scientific research trends

Dan Ofer, Hadasah Kaufman, Michal Linial*

[1]Department of Biological Chemistry, Institute of Life Sciences, The Hebrew University of Jerusalem, Jerusalem, Israel

* Correspondence author

## Abstract

Scientific research trends and interests evolve over time. The ability to identify and forecast these trends is vital for educational institutions, practitioners, investors, and funding organizations. In this study, we predict future trends in scientific publications using heterogeneous sources, including historical publication time series from PubMed, research and review articles, pre-trained language models, and patents. We demonstrate that scientific topic popularity levels and changes (trends) can be predicted five years in advance across 40 years and 125 diverse topics, including life-science concepts, biomedical, anatomy, and other science, technology, and engineering topics. Preceding publications and future patents are leading indicators for emerging scientific topics. We find the ratio of reviews to original research articles informative for identifying increasing or declining topics, with declining topics having an excess of reviews. We find that language models provide improved insights and predictions into temporal dynamics. In temporal validation, our models substantially outperform the historical baseline. Our findings suggest that similar dynamics apply across other scientific and engineering research topics. We present SciTrends, a user-friendly webtool for predicting scientific topics trends: https://hadasakaufman.shinyapps.io/SciTrend

# 1. Introduction

The progress of science relies on the discovery and dissemination of new knowledge through scientific publications. The ability to forecast future research trends is important for both researchers and organizations, as it can provide guidelines for directing scientific efforts. The importance of predicting trends over time for scientific topics is of utmost importance on funding agencies, the planning of new research centers, and more. In academia, and especially in experimental fields such as life science, it has a direct impact on faculty recruitment, curriculum, developing careers, and infrastructure investment.

The number of scientific papers published has been accelerating for at least four decades, and citation and annotation behaviors have changed with it [1]. Changes might be attributed to the continuous increase in research institutes and researchers, the increasing impact of publications for funding agencies, and academic careers [2, 3]. In addition, acceleration in publication may reflect better automated data indexing (e.g., Science Citation Index (SCI) [4]), and the establishment of a keyword annotation scheme (e.g., the medical subject headings system, MeSH [5]). Other changes within the last four decades include the expansion of open access policies, increased research originating in industry, and the ongoing increase in the total number of researchers and expected research productivity output in many fields. While presenting the current state can be based on a historical view of analytical methods, predicting future trends is far more challenging [6, 7].

Some fields, such as methods in computer science and biotechnology, have fast dynamics due to impactful technological breakthroughs (e.g., machine learning, CRISPR gene editing) [8]. In other domains, such as medicine, topics are often less dynamic over time (e.g., cancer). A topic's popularity is also influenced by social factors bias [9]. Statistical machine learning models are commonly used in attempting to forecast future events and trends [10, 11].

Traditionally, most research on scientific publication behavior focused on descriptive analyses of past trends [12], or citation networks [13, 14]. Such studies aimed to detect existing trends, such as defining topics with increasing popularity. Alternatively, predictive or prescriptive analytics approaches aim to predict future behaviors in order to answer questions such as which topics will become more (or less) popular or what can make a topic popular [15-17]. In this study, we aim to predict the future behavior of scientific topics. Note that "popular" is not referring to the number of researchers involved or to an absolute measure of the size of the scientific active community. Here, popularity refers to the relative fraction of all publications focused on a specific topic. Furthermore, changes in popularity address the time-dependent relative popularity topic compared to its own past history, as directly comparing popularity between different domains is problematic without context due to widely varying base levels of popularity.

Only a limited number of studies have investigated time-series data for forecasting future topic citations. Noorden et al. analyzed the absolute number of citations for a specific paper [18]. Other studies focused mainly on narrow domains (e.g., predicting domain-specific conferences [19, 20]). Tattershall et al. looked at binary trends within 5 years in computer science terms without continuous fine-grained prediction of the target or exogenous variables [21]. Studies analyzing the relationships between patents and research publications [22, 23] suggested that these types of publications carry mutual information.

Our goal is to predict the future popularity of diverse scientific topics, with an emphasis on life science and biomedical domains. We developed a methodology that accounts for the overall increase in the number of publications over time. We discuss novel exogenous factors, such as patents and per-domain publication trends, that can indicate if and how a field's popularity is going to change. We show meaningful results when predicting topic popularity five years into the future and discuss the potential impact of such predictions and predictive insights.

## 2. Methods

**2.1 Data compilation**

We constructed a diverse list of topics focused mostly on life science. We included neuroanatomical regions, experimental methods, and emerging biomedical technology. We also included domain-expert selections for science, technology, and engineering research topics. For example, ultrasound was selected as a technology with emerging use in broad medical diagnostics. Topics such as opioids had clear past historical and funding trends. Other topics were derived from Wikipedia's biomedical "emerging technologies" page. All together, we present 125 topics featured in PubMed publications (Supplemental **Table S1**). The counts of topical publications are based on PubMed. The PubMed database contains over 35 million published scientific works in biomedicine, life sciences, and related fields.

**2.2 Normalized publication measurement – "Popularity"**

The total quantity of publications is increasing over time. For example, the NIH's Medline database recorded 274K citeable scientific works in 1979 but 774K in 2021 (www.nlm.nih.gov/bsd/medline_cit_counts_yr_pub.html). This escalation in total publications has the potential to skew the perception of a topic's popularity if measured solely by the raw number of publications, as otherwise most topics will have more publications over time due to the confounding background trend over time. This necessitates a methodology that considers and overcomes naïve quantitative comparisons.

We adapted a measure of "popularity" that normalizes per-topic publication counts relative to the total annual output. Specifically, per year, the popularity of a given topic is calculated by dividing the number of that topic's publications by the total number of citeable, indexed publications in PubMed in the corresponding year.

We multiplied the popularity score by 100,000 in all results and figures for visual ease. Historical PubMed topic popularity was extracted using PubMed by Year (https://esperr.github.io/pubmed-by-year). Search queries are parsed by NCBI's automatic term mapping algorithm, which supports for term and acronym expansion. This normalization helps control for the overall increase in publications over time and is similar to the approach used in other works on trend and popularity analysis, such as Google Trends, where popularity signifies the proportionate presence of a topic within a total volume rather than absolute quantity, thereby offering a more accurate measure of prominence over time.

**2.3 Exogenous variables**

For each topic, we calculated the relative ratio and difference in popularity due to review and non-review (i.e., original research publication) articles. The absolute quantity of publications (rather than the per-year normalized frequency) per topic was estimated by multiplying by the number of Medline publications that year. The total fraction of US publications out of all Medline-indexed publications was added per year. The number of patents per topic, per year, was acquired by searching the U.S. Patent and Trademark Office's PatentsView database (patentsview.org/download) for each topic. The topic's relative fraction of all patents that year was also calculated. The first occurrence date of each topic (starting from 1946) was included, as was the first "valid" date (defined as an occurrence with at least 4 previous occurrences in the preceding 5 years) and the time elapsed between those dates, as well as from the prediction date. We experimented with adding the popularity of each topic's associated major MeSH term. We found that this did not improve the models, so it was removed from the final model.

**2.4 Statistical modeling**

Features were derived from the above inputs with a forecast horizon of 5 years (i.e., all time-dependent features were from at least 5 years before the target time of prediction), and the target is popularity 5 years in the future. Additional derived candidate features were explored using the SparkBeyond Automated Machine Learning framework (as used in [10, 24]) and included different lags and lagged interaction features (e.g., the difference and ratio of review to research papers for a topic), aggregated time-window-based features (e.g., the historical average popularity between 5 and 10 years beforehand), as well as differencing and similar transformations of each input time-series variable.

Machine learning regression models were trained using the Catboost [25] and Scikit-learn libraries [26] with default hyperparameters. The pretrained deep language model used the sentence-transformers package [27]. The linear model is a Scikit-learn RidgeRegressionCV model. Boosting Tree is a Catboost regression tree model with target encoding of the topic as a categorical feature. Tree and embedding augmented is the Catboost model with the topic's name embedded as additional features, extracted using the deep learning all-MiniLM-L12-v2 model [28]. Other approaches, including additional scikit-learn models and pure deep learning or statistical forecasting models, were evaluated but had significantly worse results and inferior stability, so they were not used (not shown). Boosting tree models are interpretable, fast, relatively robust, and performant in most predictive tasks and were thus used as the representative model [29].

As the last four decades better represent modern trends in scientific research, we limited the training data to the years 1979–2019. These years cover the historical USPTO PatentsView database. We do use prior historical data when generating features and conducting retrospective analyses.

For modeling, we defined for each topic the relevant starting point as the first year in which it had citations in at least 4 of the past 5 years and at least 5 valid occurrences post-1978. This threshold was used to overcome observed errors in PubMed, with some topics appearing just a few times over decades, apparently due to erroneous annotations.

Model performance was evaluated using step-forward temporal cross validation, implemented in scikit-learn, over all topic time series simultaneously. The test set was defined using scikit-learn's temporal cross-validation protocol with 30 splits.

Data and implementation details are provided in Supplementary **Table S2**. The table includes each topics popularity and descriptive features over time. The code repository is available at https://github.com/ddofer/Trends.

**2.5. Application and data implementation**

Data and implementation details are provided in Supplementary **Table S2**. The table includes 4684 lines related to the 125 selected topics. Each line includes quantified information with rich, dynamic data. The code repository is available at https://github.com/ddofer/Trends. We have used USPTO search API. The code is provided in the repository. (https://github.com/ddofer/Trends/blob/main/patentsView_api_req.ipynb). Additionally, using Shiny technology, we have developed a web application called SciTrends that allows the user to view the normalized PubMed occurrences by year for the topic of interest for the years 2023 to 2028. In addition, for each of the 125 topics used to develop the prediction model, a browsing mode for real data with a 1–6-year prediction trend is available. The application is available at: https://hadasakaufman.shinyapps.io/SciTrend/.

# 3. Results

**3.1 The dynamics of scientific terms popularity levels**

**Fig. 1** presents an overview of 125 diverse topics and their breakdown into broad themes. Detailed information on the topics and their associated scientific domains is available in Supplementary **Table S1**. Obviously, this list is not exhaustive, although the collection of topics represents a wide range of emerging and established scientific topics. In this study, we analyze these 125 topics as a showcase.

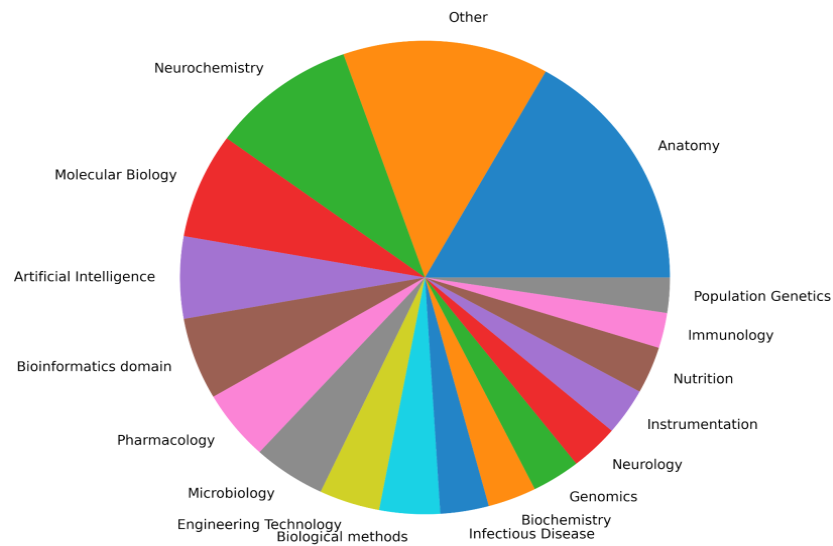

**Figure 1.** Topics by domain. Overview of the 125 topics clustered by their association with high-level field domains. A full list is available in Supplementary **Table S1**.

A sample of topics and their trend over time are illustrated in **Fig. 2A**. We follow topics' popularity (see Methods) over the past 45 years and show that topics gain, lose, and sometimes regain popularity over time. For example, while the popularity of *opioids* is quite stable, the differences over the years for *stem cells* and *neuropeptides* show very different levels of popularity and dynamics (**Fig. 2A**). There was a clear change in popularity in 1990 and 2010 for *neuropeptides* and *stem cells*, respectively.

**Fig. 2B** shows the dynamics for 62 years (1960 to 2022). Some topics display complex dynamics. For example, *RT-PCR*, which was only introduced in the early 90s, exhibits a sharp increase in popularity within 5 years (2005-2010) and a similar sharp decline that only stabilized in recent years. A similar trend was associated with *restriction enzymes*, which were the force behind molecular biology in the early days (1985–1995) and were then replaced by more simple and versatile technologies, including library preparation, CRISPR, and such. On the other hand, topics like *species conservation* and *climate change* monotonically increased in popularity, albeit only after two decades for the latter. The topic *single cell* shows a doubling in popularity that occurred in 1974, increasing popularity for two decades, and then remaining high but stable for the last 25 years.

We observed that the (normalized) mean popularity of our topics increased over time, despite our normalization. The overall change in popularity over time for all discussed topics is shown in Supplementary **Fig. S1**.

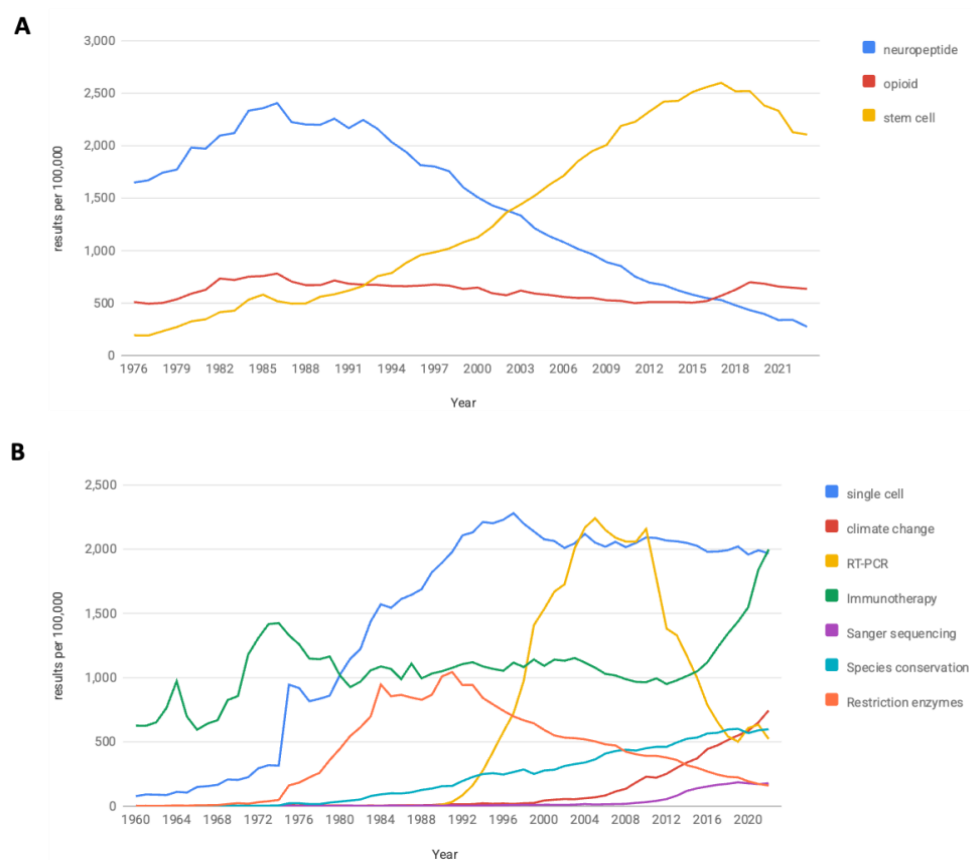

**Figure 2**. Different levels of popularity in PubMed for the indicated topics. The popularity of each topic (y-axis) is normalized per 100,000 citations out of all citations that year. **(A)** Sample of topics in the years 1976 to 2022. **(B)** Changes in the levels of popularity as extracted from PubMed covering the years 1960 to 2022. Source: PubMed by Year (see Methods).

## 3.2 Correlations with the popularity of scientific topics

Since topics' popularity includes reviews, the popularity of just review articles on a topic is significantly correlated with its overall popularity. Specifically, the total popularity of all publications (combining review and non-review articles) in that same year shows a high Pearson correlation coefficient (r=0.87) compared to 5 years before (r=0.85), as is expected. We list the ratios of review articles to research works for each topic and year (Supplementary **Table S2**). Note that it varies greatly by topic and the ongoing dynamics of a topic's popularity (Supplementary **Table S2**). Overall, a popular topic will have a higher fraction of reviews, while a rapidly growing topic may have fewer review articles relative to original research works. We conclude that the relation between the feature and its trend (i.e., changes in its popularity in the future) is different from the static popularity at the same point in time. A popular topic may have many review articles, but a growing topic may have relatively fewer reviews.

We further anticipated that patents would reflect the popularity of the topic. Indeed, overall popularity is correlated with the number of patents in that year with a Pearson rank correlation (r=0.37) and 5 years before (r=0.40). This suggests that patents may be a leading indicator for scientific publication trends, with patents preceding "valuable" research. This observation is in line with the regulation of patent applications, where confidentiality is requested to prevent its public disclosure in scientific publications before its acceptance.

We observe that topics' popularity over time is highly correlated with their own past values, e.g., from 5 years beforehand, as we might expect (r=0.953, $R^2$=0.969).

**Fig. 3** displays a diverse set of topics that we analyzed within a narrower timeframe (1990-2020). We show that popularity can change sharply with 4-5 fold increases within 5 years (e.g., *cannabidiol*, **Fig. 3A**). Many such cases represent breakouts for topics with relatively low (<50 per 100K normalized citations) previous popularity. Popular topics may also exhibit high growth and overall popularity for decades (e.g., 8000 for *ultrasound*, **Fig. 3B**). Classical topics like *vaccines* and *lipids* display rather stable dynamics for decades (**Fig. 3C**). We emphasize that some topics are quite general (e.g., *DNA*, *RNA*) and represent many subtopics that often exhibit distinctive popularity dynamics. We illustrate it by partitioning of the term *RNA* with respect to more specific related subtopics (**Fig. 3D**). Although RNA in general shows a stable trend (**Fig. 3C**), *CRISPR* (based on gRNA) and long noncoding RNA *(lncRNA)* have seen a 10-fold increase in popularity within 10 years, while transposons and ribozymes reached their maximum popularity in 2000. An interesting example refers to the 5-fold monotonic increase in rRNA (**Fig. 1D**). It is likely reflecting the emerging fields of microbiome and metagenomics, in which rRNA is used for sample characterization and microbial species identification. All these informative changes are deemed necessary by considering RNA as a general term. The normalized historical trends by years based on the PubMed database for each of the 125 topics are available in the SciTrends application (see Methods, https://hadasakaufman.shinyapps.io/SciTrend/).

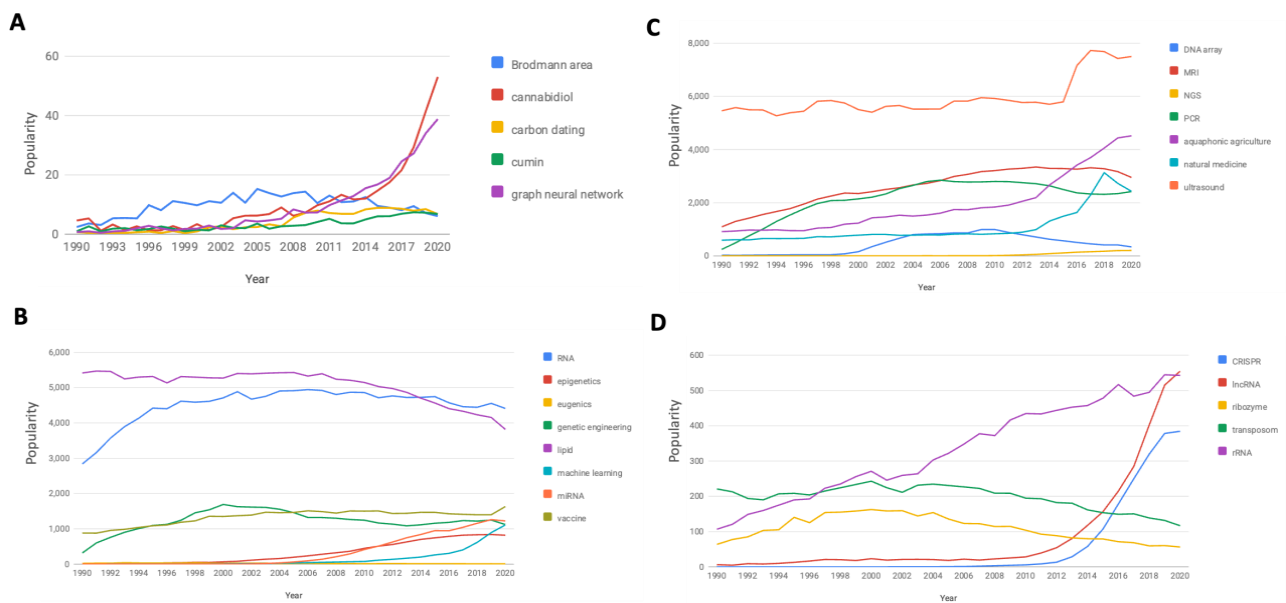

**Figure 3.** Examples from the dataset showing different levels of popularity and dynamics over three decades (1990-2020) for selected representative topics. **A-D** cover a wide range in popularity, from 50 to 8000. **(D)** Subtopics in the molecular aspects of RNAs. Note that popularity is scaled by 100,000. Image source: PubMed by year (see Methods).

**3.3 Natural language as a clue to topics' dynamics**

As shown in **Fig. 3**, the topics we discuss cover not just different scientific domains but also vary in how general or specific they are, as well as representing different concepts. Specifically, there are topics that concern methods such as *NGS* (next generation sequencing) or *PCR* (polymerase chain reaction), while others cover biological fields that are precisely defined, such as anatomical regions. We expect different fields and entities to show different behaviors and temporal dynamics. Furthermore, topics greatly vary by the size and nature of their community, e.g., medical clinicians versus ethical researchers.

We would like to incorporate relevant information into our models to enhance the potential for learning latent dynamics. For example, we anticipate that technologies might become obsolete or be replaced by modern ones (e.g., the methodologies for transgenic mice or creating cell lines). Nevertheless, terms for well-defined anatomical entities such as the *hippocampus* are unlikely to be superseded, even if interest in them varies, and the topic will remain despite being subjected to new research technology. For example, *fMRI* (functional magnetic resonance imaging) is a leading technology associated with neuroscience and cognitive psychology that exhibits massive growth in popularity thanks to its ability to allow psychology experiments with live subjects in real time. But a topic such as genetic engineering will most likely not match the dynamics of the term eugenics. The latter has a negative historical context and thus is not expected to correlate well with genetic engineering, despite their shared roots and semantic similarity.

We used a pretrained deep learning language model, all-MiniLM-L12-v2 [28] to extract a quantitative representation (embeddings) for each topic, using the name of the topic as input, and used this embedding as an additional input feature (see Tree & embedding augmented **Table 1**). The language model was trained in advance on a general text corpus and was not fine-tuned. Information can be efficiently represented using such representations [30, 31].

**Table 1** shows the model's forecasting performance. The combined tree and embedding models have the best explained variance ($R^2$) scores and overall results. All results are reported for the test set (see Methods). Lag baseline is the last valid value of the target, or the transformed target, from 5 years beforehand in a regularized linear regression model.

**3.4 Predictive model results**

Machine learning models (linear regression and boosting tree (Catboost)) were trained on different targets: (i) the popularity of each topic in 5 years ($\hat{y}$), and (ii) the percent change in a topic's popularity in 5 years relative to the present ($\Delta\hat{y}$). The latter target is a more challenging one that also implicitly neutralizes the naïve lagging baseline and reduces the bias due to differing mean levels of popularity between topics. We found that non-linear boosting tree models gave the best results (**Table 1**). Our models outperform the historical lag baseline, which presents a proxy for human guesses. We further studied the features that contributed to the model. We further evaluated at the binary level: predicting if a topic would go up or down in popularity (i.e., binary prediction). At that level, we show 88% accuracy compared to a 70% baseline (most topics increased in popularity over time). These results were stable over time.

Supplementary **Fig. S2** shows the SHAP analysis (based on Shapley values) for the model trained on the percent change target using the augmented deep learning embedding features. The top features of importance to the tree and text-augmented model are listed along with their values. We found that review articles strongly contribute to the model's performance. Numerous features associated with reviews are among the features that exhibit strong SHAP values.

**Table 1.** Model results for 5-year forecasting

| Target | Model | R² coefficient | Mean absolute error | Median absolute error | RMSE (Root mean square error) |
|---|---|---|---|---|---|
| Ŷ - popularity | Lag baseline | 0.973 | 113.80 | 44.37 | 229.12 |
| | Linear model | 0.974 | 101.19 | 45.11 | 214.82 |
| | Boosting Tree (Catboost) | 0.981 | 75.58 | 26.32 | 183.67 |
| | Tree & embedding augmented | **0.991** | **58.53** | **23.22** | **132.30** |
| Δŷ - % change in popularity | Lag baseline | 0.004 | 115.42 | 83.38 | 289.98 |
| | Linear model | 0.03 | 114.83 | 66.19 | 286.22 |
| | Boosting Tree (Catboost) | 0.306 | 60.56 | 18.47 | 241.96 |
| | Tree & embedding augmented | **0.447** | **48.29** | **15.11** | **216.04** |

**3.5 Evaluation of unseen topics**

Our results implicitly assume predicting changes in known topics. To reflect this, we added an evaluation of predicting completely novel topics over the 40 years covered in the data (**Table 2**). In this setup, the train-test split is performed at the topic level rather than the time-level, using 30-fold groupwise splits while still predicting 5 years ahead. This is considerably more challenging, as it reflects the problem of "What will the popularity of a completely unknown scientific topic be like over many decades?". This framework is unrealistically challenging, as we would expect a novel topic's behavior over decades to be predictable in advance, but it can be viewed as a proxy for lower-bound performance over completely unseen topics. We observe reduced performance and, surprisingly, no clear benefit from the text embeddings. We view the temporal evaluation setup as the more relevant one.

**Table 2**. Model predictions for topic-level splits

| Target | Model | R² coefficient | Mean absolute error | Median absolute error | RMSE (Root mean square error) |
|---|---|---|---|---|---|
| | Linear Model | 0.919 | 268.25 | 187.34 | 394.51 |

| Topic-Level split $\hat{Y}$ - popularity | Boosting Tree | 0.86 | 134.45 | 33.69 | 507.23 |
| | Tree & embedding augmented | 0.748 | 198.61 | 58.37 | 697.18 |

**3.6 Predicting the rise and fall in topics' popularity**

We limited the training data up to 2019, partially due to the extreme societal changes and publication biases in the past 2 years during the COVID-19 epidemic [32]. Nevertheless, we examined the model predictions for the present time, i.e., the per-topic model predictions for 2022, using 2017 data.

**Table 3** provides a sample of topics predicted to have the greatest relative ($\Delta\hat{y}$) change, selected from the models' top results. Model predictions for 2022 were sorted by the highest absolute predicted change relative to 2017 popularity, then selected. We list topics that had declining popularity before 2017 and are predicted to continue to decline until 2022. An example of an erroneous model prediction is *influenza*, likely due to the 2020 COVID-19 epidemic knock-on effects. We further show several topics with inverse directionality (**Table 3c**). These are other topics that were increasing in popularity in 2017 relative to their popularity level in 2016, which we predicted would decline in 2022, or that were predicted to show the opposite change trend. Specifically, their popularity decreased in 2017 relative to 2016, but nevertheless, we predict their trend to reverse. The list is sorted by the success order of the predictive model in each section.

**Table 3**. Selected predictions for 2022 with information limited to the year 2017

| Predicted popularity | Topics |
|---|---|
| a) Predicted to be more popular in 2022 Popularity: (2022>2017) | miRNA, drug repurposing, nanopore, carbon nanotubes, synthetic biology, metabolome, mononucleosis, illumina, NGS, connectome, lithium, cannabidiol, natural medicine, graph neural network, biosimilar, cumin, lncRNA, CRISPR, machine learning |
| b) Predicted to be less popular in 2022 Popularity: (2022<2017) | medulla oblongata, serotonin, DNA array, norepinephrine, neuropeptide, histamine, influenza, junk DNA, pituitary gland, ancient DNA, hypothalamus, somatosensory cortex, acetylcholine, cocaine, ribozyme |

| c) Predicted to reverse direction by 2022, relative to the 2016 to 2017 trend Popularity: (2017>2016 & 2022<2017) or (2017<2016 & 2022>2017) | eugenics, cerebellum, mononucleosis, hippocampus, MRI, antibiotic, norepinephrine, ancient viruses, zebra fish, neocortex, carbon nanotubes, carbon dating, HMM, savant |
|---|---|

Manual analysis showed most predictions to be correct, at least at the binary trend level (increasing or decreasing), with examples such as *cumin* and *graph neural networks* (**Table 3a**). We included cases where the trend for a topic (increasing or decreasing popularity) is predicted to reverse, as was indeed the observed case (e.g., *MRI, antibiotics*, **Table 3c**). **Fig. 4** shows the actual changes in popularity for 2022 based on PubMed with a resolution of 1-year, 3-years, and 5-years for selected topics from **Table 3.**

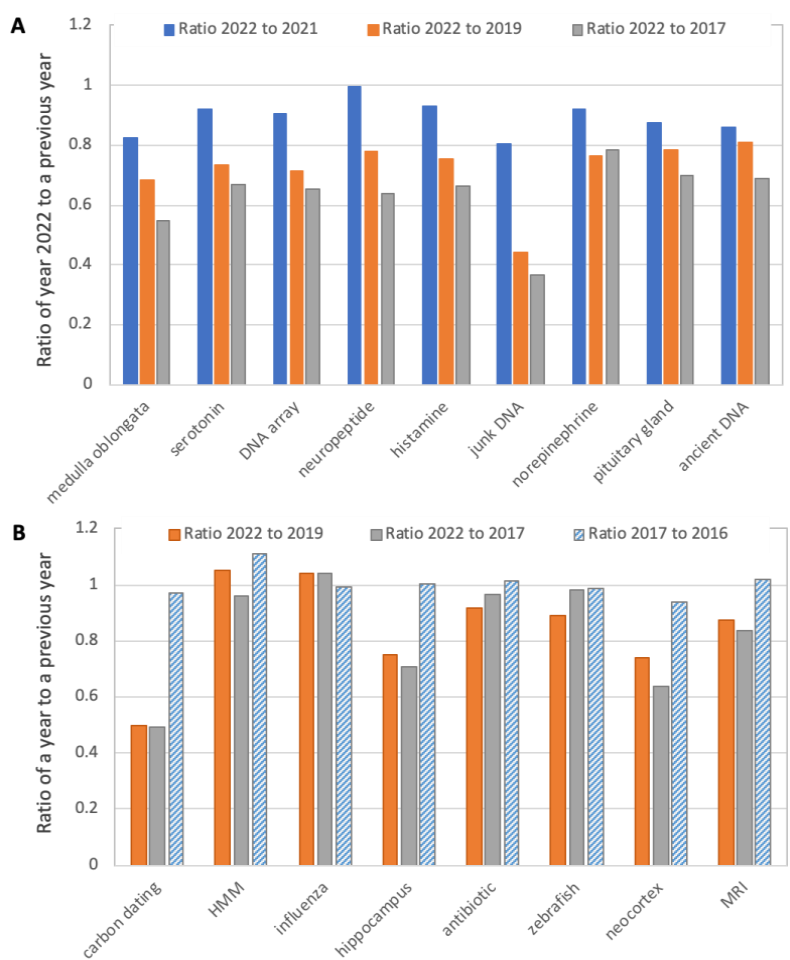

**Figure 4.** Actual changes in popularity for 2022 based on PubMed. **(A)** 2022 predictions for a decline in popularity with the trend of 1-year, 3-years, and 5-years. The topics are representatives from **Table 3b**. **(B)** Predictions for 2022 that do not agree with the previous year's trend. The change in popularity for 3-years, 5-years, and 2017 to 2016 (stripped bar) is shown. The topics are representatives from **Table 3c.**

To improve the generality of our study, we developed an application for using PubMed's current information (as of 2022), allowing the user to activate the ML model of any topic or term of interest. **Fig. 5** shows a screenshot of the results for four topics. The trend for ncRNA predicts a rather steady occurrence from 2023-2026, the prediction for APOe and Parkinson shows a steady increase; and covid19 which appeared in very high numbers (>50k of the normalized value) in 2021 and 2022, is predicted to sharply decline from 2024–2028. In addition, the SciTrends application provides a browsing option displaying the real data and the prediction trend by year for all 125 topics used for the training of the model.

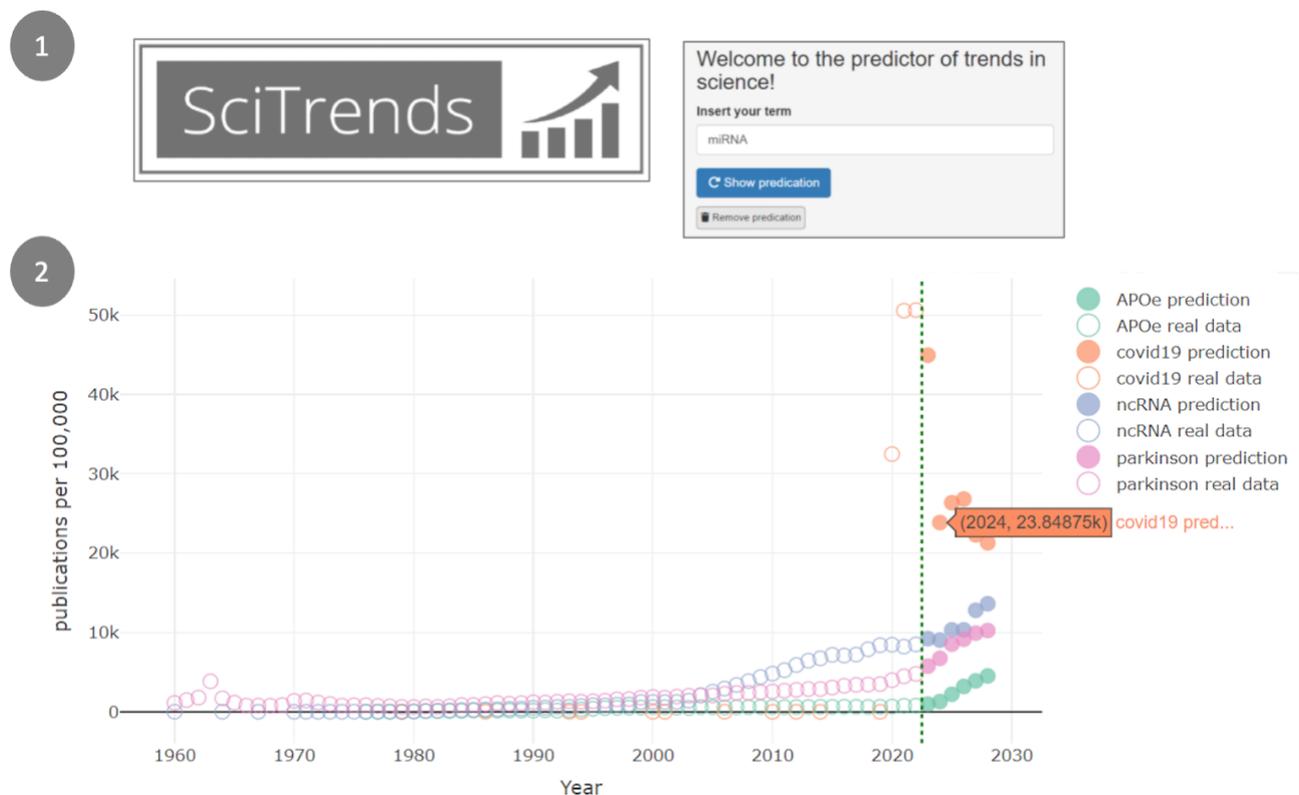

**Figure 5.** A screenshot of the SciTrends application. The user can upload any topic or term that appears in PubMed (up to 10 topics can be loaded in a single run). The split between the normalized real data and the prediction trend is marked in 2022 (a vertical dashed line). The predicted trend is shown by the filled-in colored symbol for six years from 2023 to 2028. The actual year and normalized values are shown in a popup rectangle (orange).

## 5. Discussion

While bibliometric publication trend statistics are not a perfect approximation for scientific research, they can provide valuable insights into trends in a field and help researchers and investors make informed predictions about the future direction of research. It is possible to identify emerging trends in a field and predict, to some extent, how trends may evolve over time. With respect to the known quote from Niels Bohr (1970) "It is difficult to make predictions, especially about the future", we aim to predict scientific topics in 6 years in the dynamic life science and technology fields.

Often in experimental science, such as biology, new concepts and methods precede a vast increase in related research. A classic example is the use of a microscope, which established the germ theory and preceded a vast increase in research in human health, microbiology, vaccines, and more. A similar lag can be attributed to the discovery of DNA structure, which led to the evolution of molecular biology and its key technologies (*RT-PCR*, *CRISPR*, and *NGS*). We show that it is possible to outperform the naïve baselines by which what is popular today will be popular in the coming years. Long-term forecasting is non-trivial and has been thoroughly discussed in the electronic media [33].

We present a non-exhaustive list of 125 topics, representing a wide range of emerging and established scientific and biological topics. We then searched PubMed, using the data from PubMed by Year [34] and used it to identify trends in scientific research over decades. We observed that the mean popularity of all 125 topics increased over time, despite our year-based normalization (Supplementary **Fig. S1)**. We hypothesize this might be due to improved automated annotation methods resulting in more keywords being annotated across all studies. This pattern persists even with further attempts at naively detrending the popularity target (not shown) or the percent change target ($\Delta \hat{y}$). Additionally, it might be explained as an outcome of survival bias. Specifically, we chose topics that are valid and active in the present time and most likely ignored the topics that became completely obscure and disappeared over time. Such bias may cause a shift toward more popular topics.

We used the data to train forecasting models to predict topics' popularity six years in advance. We show that our models are capable of predicting future trends in existing topics and outperforming the historical (lag) baseline. Numerous studies also used quantitative approaches while analyzing publications and citations per topic with or without multi-year dynamics [35-38]. Additionally, studies used publications to identify hidden connections between topics [39, 40]. A systematic identification of patterns in such large datasets can accelerate innovation and augment creativity [41, 42]. In tracing the evolution of topics, abstracts were parsed by rhetorical framing [43]. They found that topics referred to in results sections tend to decline, while the opposite holds for topics appearing in methods sections. Such findings are in accordance with our study, where the abundance of reviews relative to original research articles often precedes stagnating or declining topic popularity.

An important concept in forecasting tasks are leading indicators, which are exogenous variables that provide predictive, potentially causal information about targets of interest. Patents are an intriguing candidate for predicting similar turning points in scientific research [23]. Commercially relevant works are advised to obtain defensive patents prior to publication. In this case, patents should most likely precede publications. We examine the CRISPR-based gene editing technology as a case study [44]. We observe from the data that the number of patents in earlier years is a strong, leading indicator and better predicts research papers than the number of patents in that same year (Supplementary **Table S3**). Recall that, typically, temporal distance reduces the correlation between variables. The Pearson correlation between the number of CRISPR publications and patents filed the same year is r = 0.93, while the correlation with patents filed 1 year before is r = 0.98. Using patent values from 1 year into the future gives an even lower correlation with popularity (r = 0.88). This is in line with the hypothesis that patents might be a leading indicator for research publications. This observation is also in line with the regulation of patent applications where confidentiality is requested to prevent its public disclosure in scientific publications before its acceptance.

In this study, we found that a relatively high number of reviews correlates with reduced future topic popularity. This could be because review articles are often published when a field or a technology is mature. When a field is in its early stages, a lot of new research is published, especially in experimental sciences. In other instances, such as the recent COVID-19 global pandemic, the field is so dynamic that the relative number of reviews was suppressed in view of the flood of original publications. The 3-year period of COVID-19 also affected peer-reviewed protocols, the time lag in publications, and the abrupt change in science that was made in the fields that are related to public health, epidemiology, virology, and vaccination. It is likely that once a field matures, both conceptual and literature surveys will be presented, and it may stagnate. This is supported by different works on the relative locations and context of topics in articles. Examples include MRI and fMRI in cognitive and neuroscience research, or deep learning in computer vision [45], language, and biology [7, 16, 46].

This research is subject to several limitations. We have used PubMed as the ground-truth source for publication counts. However, PubMed does not include records from the new channels for publishing in science, such as conference proceedings, open archives (e.g., BioRxiv), or online collections. Moreover, PubMed focuses mainly on medicine and biomedical sciences, and coverage of other research topics is limited. The generality of our predicting model was not yet tested on other publication resources, such as scientific blogs, or domains that are not generally published on PubMed, such as social sciences or linguistics.

A cautionary aspect is cases of semantic transitions in terminology switching, where a topic may be referred to using different nomenclatures over time, e.g., NGS to deep sequencing. The PubMed search algorithms we rely on help mitigate this by handling many acronyms automatically (e.g., "DNA" and "deoxyribonucleic acid"), but this is an aspect that our framework does not directly handle.

Defining something as popular is fuzzy and contextual. It can be viewed in relation to itself, i.e., a relative increase or decrease in popularity compared to the past (for example, neuropeptides, influenza, and mRNA vaccines), or in relation to other topics. For example, cancer is extremely "popular", as it is a widely used term in immunology, cell biology, computational biology, medicine, and other scientific fields. In contrast, mentioning the genes that drive cancer (e.g., TP53, BRCA1) is far less popular.

The trends in scientific and technological topics are influenced by confounding effects ranging from human curiosity (e.g., space science), media coverage, and public awareness (e.g., gene therapy), or pressing challenges in public health (e.g., vaccines, climate change). These factors shape public and institutional interest in specific areas of science and technology. Thus, applications of this methodology to predict popularity as the basis for planning must also take human social context into account. For example, to help predict if something is a temporary "flash in the pan" or a meaningful, breakout topic that will remain relevant for years to come.

We found that unstructured text embeddings of topics using just their names provided additional information. The best predictive models, out of those tested, likely reflect the models' added capacity to learn latent dynamics between domains [7, 24]. However, these results should be viewed with caution, as the underlying language model was trained on data from 2018. While it was not exposed to our tasks, the possibility of future information leakage cannot be discounted, barring a full retraining of the model for every year covered. We leave this as a direction for future research, having established the value of such information for the forecasting task.

## 6. Conclusions

The dynamics and value ranges of the topics vary greatly by topic and time and may span over 4 orders of magnitude, even after normalization. Despite these dynamic and scale-varying targets, our results suggest that scientific publication trends are predictable years in advance using historical data as well as patents and in-domain publication trends, such as the number of reviews relative to research articles. We suggest that such methods can be of great benefit for planning critical decisions regarding career development, technological implantation, training, and education, as well as for early-stage researchers investing in infrastructure and training. To empower the utility of our prediction models, we developed SciTrends as an interactive application that presents the profile of any topic of interest covered by PubMed and its trend prediction for the following 6 years.

**Author contribution statement**

Dan Ofer designed the study and analyzed the data. Hadasa Kaufman developed the SciTrends application. Dan Ofer and Michal Linial carried out interpretation and visualization. All coauthors wrote and revised the manuscript.

**Data availability**

Data and implementation details are provided in Supplementary **Table S2**. The code repository is available in https://github.com/ddofer/Trends. SciTrends application is available at: https://hadasakaufman.shinyapps.io/SciTrend/

**Declaration of competing interest**

The authors declare that they have no competing financial interests or personal relationships that could have appeared to influence the work reported in this paper.

**Acknowledgments**

We thank the members from D. Shahaf and M. Linial laboratories for sharing their ideas and valuable discussions. We thank N. Rappoport and R. Zucker for supporting the web application. This work was partially supported by the Center for Interdisciplinary Data Science Research (CIDR, #3035000440) at the Hebrew University, Jerusalem.

## 7. References

1. Bornmann, L. & Mutz, R. (2015) Growth rates of modern science: A bibliometric analysis based on the number of publications and cited references, *Journal of the Association for Information Science and Technology.* **66**, 2215-2222.
2. Fairclough, R. & Thelwall, M. (2022) Questionnaires mentioned in academic research 1996–2019: Rapid increase but declining citation impact, *Learned Publishing.* **35**, 241-252.
3. Tang, L., Shapira, P. & Youtie, J. (2015) Is there a clubbing effect underlying C hinese research citation Increases?, *Journal of the Association for Information Science and Technology.* **66**, 1923-1932.
4. Larsen, P. & Von Ins, M. (2010) The rate of growth in scientific publication and the decline in coverage provided by Science Citation Index, *Scientometrics.* **84**, 575-603.
5. Mao, Y. & Lu, Z. (2017) MeSH Now: automatic MeSH indexing at PubMed scale via learning to rank, *J Biomed Semantics.* **8**, 15.
6. Salganik, M. J. (2023) Predicting the future of society, *Nat Hum Behav*.
7. Ofer, D., Brandes, N. & Linial, M. (2021) The language of proteins: NLP, machine learning & protein sequences, *Comput Struct Biotechnol J.* **19**, 1750-1758.
8. Chu, J. S. G. & Evans, J. A. (2021) Slowed canonical progress in large fields of science, *Proc Natl Acad Sci U S A.* **118**.
9. Klein, D. B. & Stern, C. (2009) Groupthink in academia: Majoritarian departmental politics and the professional pyramid, *The Independent Review.* **13**, 585-600.
10. Cohen, S., Dagan, N., Cohen-Inger, N., Ofer, D. & Rokach, L. (2021) ICU survival prediction incorporating test-time augmentation to improve the accuracy of ensemble-based models, *IEEE Access.* **9**, 91584-91592.
11. Spiliotis, E., Assimakopoulos, V., Makridakis, S. & Assimakopoulos, V. (2022) The M5 Accuracy competition: Results, findings and conclusions, *Int J Forecast*.
12. Bornmann, L. & Haunschild, R. (2022) Empirical analysis of recent temporal dynamics of research fields: Annual publications in chemistry and related areas as an example, *Journal of Informetrics.* **16**, 101253.
13. Effendy, S. & Yap, R. H. (2017). Analysing trends in computer science research: A preliminary study using the microsoft academic graph. Paper presented at the *Proceedings of the 26th international conference on world wide web companion*.

14. Abelson, P. H. (1964) Trends in Scientific Research: Rapid evolution of the frontiers is a hazard for scientists young and old, *Science.* **143**, 218-223.

15. Serrano Najera, G., Narganes Carlon, D. & Crowther, D. J. (2021) TrendyGenes, a computational pipeline for the detection of literature trends in academia and drug discovery, *Sci Rep.* **11**, 15747.

16. Savov, P., Jatowt, A. & Nielek, R. (2020) Identifying breakthrough scientific papers, *Information Processing & Management.* **57**, 102168.

17. Mazov, N., Gureev, V. & Glinskikh, V. (2020) The methodological basis of defining research trends and fronts, *Scientific and Technical Information Processing.* **47**, 221-231.

18. Van Noorden, R. (2013) Formula predicts research papers' future citations, *News, Nature.* **3**.

19. Nezhad, F. G., Osareh, F. & Ghane, M. R. (2022) Forecasting the Subject Trend of International Library and Information Science Research by 2030 Using the Deep Learning Approach, *International Journal of Information Science & Management.* **20**.

20. Abuhay, T. M., Nigatie, Y. G. & Kovalchuk, S. V. (2018) Towards predicting trend of scientific research topics using topic modeling, *Procedia Computer Science.* **136**, 304-310.

21. Tattershall, E., Nenadic, G. & Stevens, R. D. (2020) Detecting bursty terms in computer science research, *Scientometrics.* **122**, 681-699.

22. Park, M., Leahey, E. & Funk, R. J. (2023) Papers and patents are becoming less disruptive over time, *Nature.* **613**, 138-144.

23. Griffin, T. D., Boyer, S. K. & Councill, I. G. (2010). Annotating patents with Medline MeSH codes via citation mapping. Paper presented at the *Advances in Computational Biology*.

24. Ofer, D. & Linial, M. (2022) Inferring microRNA regulation: A proteome perspective, *Frontiers in Molecular Biosciences.* **9**, 989.

25. Dorogush, A. V., Ershov, V. & Gulin, A. (2018) CatBoost: gradient boosting with categorical features support, *arXiv preprint arXiv:181011363*.

26. Pedregosa, F., Varoquaux, G., Gramfort, A., Michel, V., Thirion, B., Grisel, O., Blondel, M., Prettenhofer, P., Weiss, R. & Dubourg, V. (2011) Scikit-learn: Machine learning in Python, *the Journal of machine Learning research.* **12**, 2825-2830.

27. Reimers, N. & Gurevych, I. (2019) Sentence-bert: Sentence embeddings using siamese bert-networks, *arXiv preprint arXiv:190810084*.

28. Wang, W., Bao, H., Huang, S., Dong, L. & Wei, F. (2020) Minilmv2: Multi-head self-attention relation distillation for compressing pretrained transformers, *arXiv preprint arXiv:201215828*.

29. Zhang, C., Liu, C., Zhang, X. & Almpanidis, G. (2017) An up-to-date comparison of state-of-the-art classification algorithms, *Expert Systems with Applications.* **82**, 128-150.

30. Ofer, D. & Shahaf, D. (2022) Cards against AI: Predicting humor in a fill-in-the-blank party game, *arXiv preprint arXiv:221013016*.

31. Brandes, N., Ofer, D., Peleg, Y., Rappoport, N. & Linial, M. (2022) ProteinBERT: a universal deep-learning model of protein sequence and function, *Bioinformatics.* **38**, 2102-2110.

32. Ahmad, T., Murad, M. A., Baig, M. & Hui, J. (2021) Research trends in COVID-19 vaccine: a bibliometric analysis, *Hum Vaccin Immunother.* **17**, 2367-2372.

33. Forecasting, C. (2023) Insights into the accuracy of social scientists' forecasts of societal change, *Nat Hum Behav*.

34. Sperr, E. V. (2006) Libraries and the future of scholarly communication, *Molecular cancer.* **5**, 1-2.

35. Braam, R. R., Moed, H. F. & Van Raan, A. F. (1991) Mapping of science by combined co-citation and word analysis. I. Structural aspects, *Journal of the American Society for information science.* **42**, 233-251.

36. Tahamtan, I. & Bornmann, L. (2019) What do citation counts measure? An updated review of studies on citations in scientific documents published between 2006 and 2018, *Scientometrics.* **121**, 1635-1684.

37. Bornmann, L. (2013) The problem of citation impact assessments for recent publication years in institutional evaluations, *Journal of Informetrics.* **7**, 722-729.

38. Voytek, J. B. & Voytek, B. (2012) Automated cognome construction and semi-automated hypothesis generation, *Journal of neuroscience methods.* **208**, 92-100.

39. Marx, W. & Bornmann, L. (2016) Change of perspective: bibliometrics from the point of view of cited references—a literature overview on approaches to the evaluation of cited references in bibliometrics, *Scientometrics.* **109**, 1397-1415.

40. Kleminski, R., Kazienko, P. & Kajdanowicz, T. (2022) Analysis of direct citation, co-citation and bibliographic coupling in scientific topic identification, *Journal of Information Science.* **48**, 349-373.

41. Hope, T., Chan, J., Kittur, A. & Shahaf, D. (2017). Accelerating innovation through analogy mining. Paper presented at the *Proceedings of the 23rd ACM SIGKDD International Conference on Knowledge Discovery and Data Mining*.

42. Kittur, A., Yu, L., Hope, T., Chan, J., Lifshitz-Assaf, H., Gilon, K., Ng, F., Kraut, R. E. & Shahaf, D. (2019) Scaling up analogical innovation with crowds and AI, *Proceedings of the National Academy of Sciences.* **116**, 1870-1877.

43. Prabhakaran, V., Hamilton, W. L., McFarland, D. & Jurafsky, D. (2016). Predicting the rise and fall of scientific topics from trends in their rhetorical framing. Paper presented at the *Proceedings of the 54th Annual Meeting of the Association for Computational Linguistics (Volume 1: Long Papers)*.

44. Sander, J. D. & Joung, J. K. (2014) CRISPR-Cas systems for editing, regulating and targeting genomes, *Nat Biotechnol.* **32**, 347-55.

45. Krizhevsky, A., Sutskever, I. & Hinton, G. E. (2017) Imagenet classification with deep convolutional neural networks, *Communications of the ACM.* **60**, 84-90.

46. Devlin, J., Chang, M.-W., Lee, K. & Toutanova, K. (2018) Bert: Pre-training of deep bidirectional transformers for language understanding, *arXiv preprint arXiv:181004805*.

## Supplemental Tables

**Table S1**. The list of 125 topics and their associated domain (.xls)

**Table S2.** Topic list along with dynamic information (.xls)

**Table S3.** CRISPR patents over time (.xls)

## Supplemental Figures

**Figure S1**
**Figure S2**

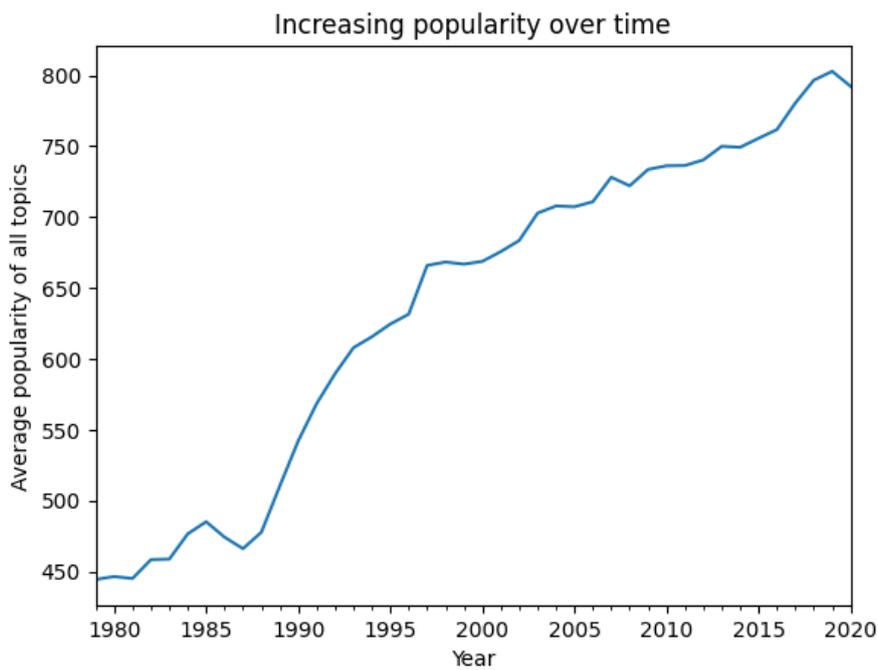

**Figure S1.** Change in popularity over time for all discussed topics

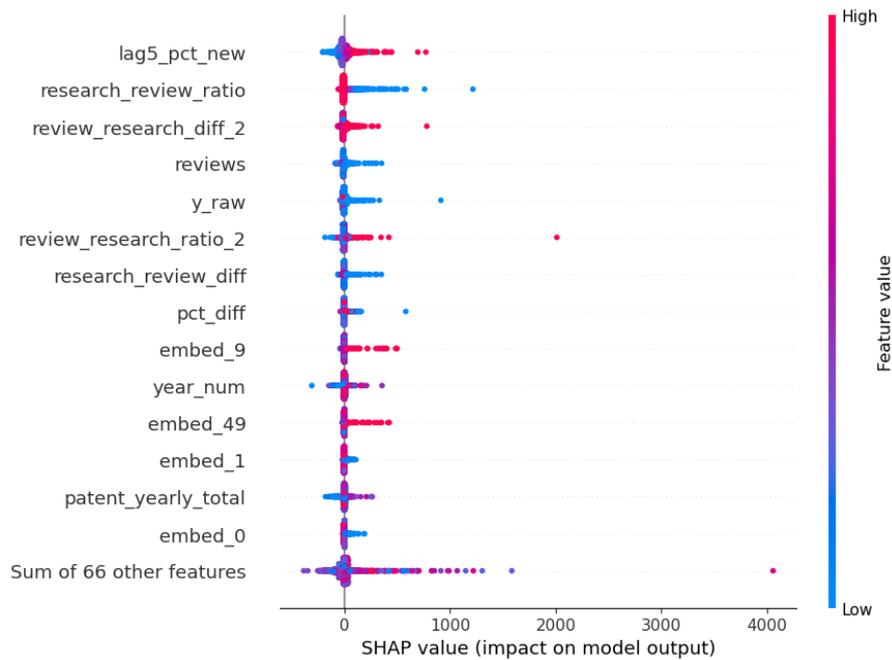

**Figure S2** - Feature importance to the tree and text augmented model. The figure shows Shapley values for the model trained on the percent change target, using the augmented deep learning embedding features.

Briefly, Embed_X features are component vectors of the topic embedding from the deep learning text embedding model. Research_review_ratio is the ratio of research to review publications. Review_research_diff is the difference (rather than ratio, i.e., reviews-popularity minus research-popularity, for the topic). Lag5_pct_new is the percent change in a topic's popularity relative to 5 years before. Pct_diff is the percent change in a topic's popularity relative to 1 year before. Y_raw is the undifferenced value of the target. Year_num is the year (time elapsed) of prediction. Patent_yearly_total is the sum total patents released that year, relevant to the topic. Features are explained in detail in the code.